# Comprehensive reevaluation of acetaldehyde chemistry and the underlying uncertainties

Xinrui REN[a], Hongqing WU[a], Ruoyue TANG[a], Yanqing CUI[a], Mingrui WANG[a], Song CHENG[a,b*]

[a] *Department of Mechanical Engineering, The Hong Kong Polytechnic University, Kowloon 999077, Hong Kong*
[b] *Research Institute for Smart Energy, The Hong Kong Polytechnic University, Kowloon 999077, Hong Kong*

___

**Abstract**

Understanding the combustion chemistry of acetaldehyde is crucial to developing robust and accurate combustion chemistry models for practical fuels, especially for biofuels. This study aims to reevaluate the important rate and thermodynamic parameters for acetaldehyde combustion chemistry. The rate parameters of 79 key reactions are reevaluated using >100,000 direct experiments and quantum chemistry computations from >900 studies, and the thermochemistry ($\Delta h_f$(298K), $s^0$(298K) and $c_P$) of 24 key species are reevaluated based on the ATCT database, the NIST Chemistry WebBook, the TMTD database, and 35 published chemistry models. The updated parameters are incorporated into a recent acetaldehyde chemistry model, which is further assessed against available fundamental experiments (123 ignition delay times and 385 species concentrations) and existing chemistry models, with clearly better performance obtained in the high-temperature regime. Sensitivity and flux analyses further highlight the insufficiencies of previous models in representing the key pathways, particularly the branching ratios of acetaldehyde- and formaldehyde-consuming pathways. Temperature-dependent and temperature-independent uncertainties are statistically evaluated for kinetic and thermochemical parameters, respectively, where the large differences between the updated and the original model parameters reveal the necessity of reassessment of kinetic and thermochemical parameters completely based on direct experiments and theoretical calculations for rate and thermodynamic parameters. Based on extensive data calculations, the important rate and thermodynamic parameters for acetaldehyde chemistry are reevaluated and their uncertainties are determined. Further uncertainty quantification and optimization can be conducted to improve the acetaldehyde chemistry model, which is currently under progress in the authors' group.

*Corresponding authors: Song Cheng

Phone: +852 2766 6668

Email:songcheng@polyu.edu.hk

**Highlights**

- This study has reevaluated acetaldehyde chemistry (including the rate parameters of 79 reactions and the thermodynamic parameters of 24 species) purely based on available physical information (i.e., direct experiments and quantum chemistry computations), including over 100,000 direct experiments and quantum chemistry computations from >900 studies.
- An updated chemical model for acetaldehyde was developed that demonstrated better agreement with all available indirect experiments from fundamental reactors.
- Moreover, the study has defined and conditioned the temperature-dependent uncertainties for the all important rate parameters and the temperature-independent uncertainties for all the important species thermodynamic parameters, which are statistical representations of the model uncertainty over the re-evaluated parameter values. This uncertainty domain will enable real-time optimization of large-scale chemistry models with minimized possibility of encountering data inconsistency.

# 1. Introduction

Acetaldehyde is an important intermediate and a pollutant species during the combustion of fossil fuels and biofuels. In addition, acetaldehyde chemistry marks the final steps of combustion chemistry of $C_2$ and heavier hydrocarbons [1] (particularly alcohols [2,3]), and combustion chemistry models for these hydrocarbons have been constructed based on the acetaldehyde chemistry models. Understanding the combustion chemistry of acetaldehyde is crucial to developing robust and accurate combustion chemistry models for practical fuels, especially for biofuels.

The oxidation of acetaldehyde chemistry has been extensively studied in the past. The earliest article on acetaldehyde combustion can be dated back to 1934, where Rice and Herzfeld [4] first proposed and analyzed a reaction set for acetaldehyde decomposition. Following this, Beeley et al. [5] devised a high-temperature reaction mechanism for acetaldehyde and compared the modeling results with ignition delays obtained in a shock tube at equivalence ratios 0.5-3.0 and temperatures from 1550 to 1850K. Later, Kaiser et al. [6] carried out acetaldehyde oxidation experiments using a low-pressure static reactor over the low-temperature range of 553-713K. A mechanism of acetaldehyde was constructed which was able to fairly predict experimental features and quantitative product concentrations. Borisov et al. [7] developed a mechanism of acetaldehyde oxidation at intermediate and high temperatures, which was subsequently validated through a series of experiments.

Apart from the early studies as mentioned above, more experimental efforts have been addressed to understand acetaldehyde combustion chemistry since 2000. Dagaut et al. [8] studied the oxidation of acetaldehyde in a jet-stirred reactor (JSR) at temperatures of 900-1300 K and pressures of 1-10 atm. Ignition delay times in a shock tube were also measured in a wide range of conditions ($0.5 \leq \varphi \leq 2$, 1230-2530 K, 2-5 atm). Won et al. [9] measured the ignition delay of acetaldehyde behind a reflected shock wave at high temperatures, i.e., 1320-1897 K, at 100 torr, and a mechanism consisting of 34 species and 110 reactions was proposed. Mével et al. [10] also studied high-temperature pyrolysis and oxidation of acetaldehyde in a shock tube at 1295–1580 K and 306-404 KPa. They excluded reactions related to low-temperature oxidation and model the measured ignition delay times of acetaldehyde with reasonable agreement achieved. Tao et al. [11] measured the chemical structures of low-pressure laminar premixed acetaldehyde flames at equivalence ratios of 1.7 and 1.0. They identified and quantified about 40 species in the post-flame zone, which provided more detailed information on the high-temperature chemistry of acetaldehyde combustion. Recently, Zhang et al. [12] focused on the chain-branching reactions which affected the low-temperature oxidation rate of acetaldehyde and developed a kinetic model of acetaldehyde. Low-temperature JSR oxidation of acetaldehyde was also carried out at a range of thermodynamic and fuel loading conditions ($0.5 \leq \varphi \leq 4$, 460-900K, 710-720 Torr), and a variety of reactive species were identified and quantified. They found that newly added reactions, such as H-atom abstraction from acetaldehyde by methyl peroxyl radical, had great impact on modeling results. Additionally, Tao et al. [13] also developed a new kinetic model for the oxidation of acetaldehyde recently, based on their previous study [14]. They reported the mole fraction profile of 31 species in a counterflow flame at 600 Torr and ignition delay times at 10 atm and 700-1100K in a rapid compression machine (RCM), and validated the new kinetic model using these experiments.

Despite the studies as reviewed above, the relevant studies on acetaldehyde fundamental combustion are still lacking, particularly those at high pressure conditions that are more relevant to advanced propulsion systems. As such, acetaldehyde chemistry has been somewhat overlooked and ill-conditioned in existing models for large hydrocarbons, as critically pointed out by Cheng et al. [15] in a recent study, where an ethanol chemistry model and a gasoline surrogate chemistry model agreed well with the high-pressure ignition delay times of ethanol and gasoline surrogates, respectively, while showing great errors for acetaldehyde.

On the other hand, chemistry models can be improved in a systematic manner with a reasonable amount of effort while uncertainty quantification (UQ) and optimization frameworks. Different approaches have been developed for UQ and optimization of chemical kinetic models. The underlying frameworks can be categorized as either probabilistic or deterministic, depending on how the uncertainty is treated. The most influential deterministic approach is the "bound-to-bound data collaboration" (B2B-DC) pioneered by Frenklach and co-workers [16], which prescribes fixed uncertainty bounds for involved parameters and determines the overall model uncertainty using the maximum and minimum of the simulation results. In contrast, probabilistic approaches address the uncertainty of the parameters and the model with probability density functions, such as the polynomial chaos expansion (MUM-PCE) method developed by Wang and co-workers [17], the global parameter optimization algorithm (GPOA) developed by Turányi and co-workers [18], and the ANN-MCMC method developed by Yang and co-workers [19]. Both the deterministic and probabilistic approaches can be used to optimize chemical kinetic models, known as the *reverse* problems. For instance, the use of B2B-DC to optimize a recent syngas model [20], MUM-PCE to optimize the USC Mech II [21], and GPOA to optimize a methanol model [22]. These methods have been proved very effective to improve the accuracy of chemistry models, which unfortunately have not been applied to acetaldehyde yet, despite the vast data available.

Therefore, the objectives of this study are two-fold: (a) to reevaluate the acetaldehyde chemistry purely based the available rate and thermodynamic data; and (b) to define a chemistry model and condition it uncertainty purely based on direct physical information. To this end, all the important kinetic and thermodynamic parameters are identified and reevaluated based on >100,000 direct experiments and quantum chemistry computations from >900 studies. Following this, the reevaluated kinetic and thermodynamic parameters are incorporated into an updated model, which is further validated against all available experiments and compared among various models, and analyzed via sensitivity and flux analyses. Finally, the uncertainties of the reevaluated parameters are physically characterized, which is followed by statistical analyses of model uncertainty.

## 2. Experimental and computational methods

### 2.1 Chemical kinetic model of acetaldehyde

The last version of the acetaldehyde chemistry model developed by Tao et al. [13] is adopted in this study for further analysis, while the other models, including those proposed by Zhang et al. [12], Daugaut et al. [8], and Mével et al. [10], are used for comparisons (as discussed in the following sections). The mechanism from [13] has been validated against a number of experimental and literature data, demonstrating reasonably good performance under wide ranges of temperatures (300–2300 K) and pressures (0.02–10 atm) [13]. It should be noted that using a different acetaldehyde model will have negligible influences as the present work will reevaluate all the important reactions in acetaldehyde chemistry that are almost identical in different acetaldehyde chemistry models. Additionally, our recent studies[1, 23] have found that the H-atom abstractions from acetaldehyde by $CH_3O$ and $CH_3O_2$ radicals are quite important. As such, the rate coefficients and thermochemical parameters involved in the H-atom abstraction reactions from formaldehyde and acetaldehyde by $CH_3O$ and $CH_3O_2$ radicals are updated following [1].

### 2.2 Sensitivity analysis

Brute force sensitivity analysis[24] is first conducted under the selected experimental conditions to obtain the kinetic and thermodynamic parameters that influence the model predictions. The conditions for sensitivity analysis are shown in Figure 1, which covers the typical temperature and pressure conditions of interest to the combustion community. Fuel rich conditions are not covered in Fig. 1 as the reactions identified at the selected conditions have already included the top sensitive reactions at fuel-rich conditions.

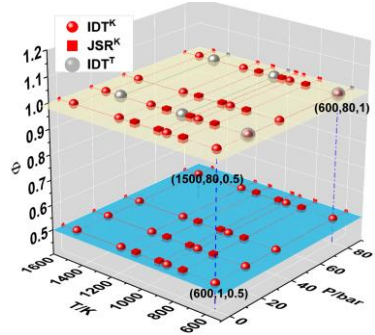

Figure 1. The conditions for sensitivity analysis for kinetic parameters on IDT (red ball), kinetic parameters on JSR (red box), and thermodynamic parameters on IDT (gray ball).

Sensitivity analyses of kinetic parameters are performed by changing the pre-exponential factors of each reaction. The sensitivity coefficients are defined as $SC_i^A = \ln(\Delta \tau / \tau) / \ln(\Delta k / k)$, where $\Delta \tau$ is the simulated result after multiplying the original rate constant by 2, i.e., $\Delta k = 2 * k$, and $\tau$ is the original result. Sensitivity analyses are not conducted at fuel-rich conditions as the reactions from these conditions have already been identified from the Sensitivity analyses stoichiometric and fuel-lean conditions. Three species are selected in the sensitivity analysis of JSR experiments, namely $CH_3CHO$, $CH_2O$, and $CO$, which are key intermediates during the initial and later oxidation stages of acetaldehyde.

Sensitivity analyses of thermodynamic parameters are conducted following the method proposed by Lehn et al. [25]. The normalized sensitivity coefficients of ignition delay time on heat capacities are calculated via Eq.1

$$SC_i^{c_p} = \frac{\partial \ln \tau_{ig}}{\partial \ln c_{p,i}} = \frac{\tau_{ig}^{mod} - \tau_{ig}^{ori}}{\tau_{ig}^{ori}} \frac{1}{SAF_i - 1} \quad (1)$$

The modified heat capacity is realized by multiplying the polynomial coefficients with the sensitivity analysis factor $SAF_i$ so that

$$a_{i,j}^{mod} = SAF_i * a_{i,j}^{ori}, j = 1,2,\dots,5 \quad (2)$$

where $SAF_i$ is the sensitivity analysis factor of species $i$. Since the sensitivity is desired with respect to heat capacity only, the coefficients $a_{i,6}$ and $a_{i,7}$ have to be adapted as well to ensure that $\Delta h^{mod}_{f,i}(298K) = \Delta h^{ori}_{f,i}(298K)$ and $s^{mod}_i(298K) = s^{ori}_i(298K)$. As such, the sensitivity coefficients for enthalpies and standard entropies are defined as

$$SC_i^h = \frac{\partial \ln \tau_{ig}}{\partial \ln (\exp \frac{\Delta h_{f,i}(298K)}{R \cdot 298K})} = \frac{\tau_{ig}^{mod} - \tau_{ig}^{ori}}{\tau_{ig}^{ori}} \frac{R \cdot 298K}{\delta \Delta h_{f,i}} \quad (3)$$

$$SC_i^s = \frac{\partial \ln \tau_{ig}}{\partial \ln (\exp \frac{s_i^0(298K)}{R})} = \frac{\tau_{ig}^{mod} - \tau_{ig}^{ori}}{\tau_{ig}^{ori}} \frac{R}{\delta s_i^0} \quad (4)$$

where $\delta \Delta h_{f,i}$ and $\delta s_i^0$ are the absolute value of $\Delta h_{f,i}(298K)$ and $s_i^0(298K)$, respectively, $a_{i,6}$ and $a_{i,7}$ are increased in the brute-force method.

The 20 most sensitive reactions and species are selected from each condition, from which a total of 79 reactions and 24 species are obtained. Results of the sensitivity analyses are reported in the SMM2&3.

*2.2 Statistical treatment of parameters*

For each sensitive reaction, direct measurements and theoretical determinations of the rate coefficients in both reverse and forward directions are collected from the NIST Chemical Kinetics Database [26] and review articles [27–32]. The reverse rates are then converted into forward rates based on chemical equilibrium calculations. All direct experiments and quantum chemistry computations within the temperature range of 300-2500K are included (a few data belonging to obsolete measurements are not considered), which are used to reevaluate the rate parameters of the specific reactions. For elementary reactions containing a third body, the low-pressure limiting rate is reevaluated. The adopted direct experiments and quantum chemistry computations, as well as the reevaluation process and the reevaluated rate parameters are summarized in SMM4. The fitting process during kinetic parameter reevaluation is briefly described herein.

Three Arrhenius parameters (namely $A$, $n$ and $Ea$) are reevaluated for all selected reactions using least squares curve fitting. In one case (R83, $HO_2+HO_2=H_2O_2+O_2$), six Arrhenius parameters (i.e., duplicated reactions) are used. The temperature dependence of rate coefficient $k$ is described by the modified Arrhenius equation $k(T) = A\{T\}^n \exp(-Ea/RT)$, and the linearized form of which is

$$\kappa(T) = \alpha + n \ln\{T\} - \varepsilon T^{-1} = \theta^T p \tag{5}$$

where $\kappa(T) = \ln\{k(T)\}$ represents the logarithm of the rate coefficient, $\alpha = \ln\{A\}$, $\varepsilon = Ea/R$, $\theta = (1, lnT, -T^{-1})^T$, and $p = (\alpha, n, \varepsilon)^T$ is the set of transformed Arrhenius parameters. The equation for the least squares fitting is defined as

$$F(T; \tilde{\mathbf{x}}) = \tilde{\alpha} + \tilde{n} \ln\{T\} - \tilde{\varepsilon} T^{-1} \tag{6}$$

where the vector $\tilde{\mathbf{x}} = (\tilde{\alpha}, \tilde{n}, \tilde{\varepsilon})$ represents the Arrhenius parameters that need to be reevaluated. The optimization objective function during the fitting process is described as

$$\Phi(x^*) = \sum_{m=1}^{N} (\kappa(T_m) - F(T; \tilde{\mathbf{x}}))^2 \tag{7}$$

where $x^*$ is the optimal parameter set that minimizes the deviation between the fitted rate and the original data; $N$ is the number of temperature conditions used for re-fitting Arrhenius parameters; $T_m$ is the actual temperature for experimental measurements and theoretical calculations, $T_m \in [300, 2500]K$; and $\kappa(T_m)$ is the logarithm of the corresponding rate coefficient, as determined from Eq. 5.

For each sensitive species, the mean value of $\Delta h_f(298K)$ is determined from reported data that are available in the Active Thermochemical Tables (ATCT) [33], the NIST Chemistry WebBook [26], and the Third Millennium Thermodynamic Database (TMTD) [34]. For four species (namely $CH_3CO_3$, $HOCH_2O$, $CH_3CO_3H$, and $O_2CH_2CHO$), no thermochemical information can be found in the above databases. On the other hand, the literature data for $s^0(298K)$ and $c_p$ are much more lacking. Therefore, $s^0(298K)$ and $c_p$ for all species, and the $\Delta h_f(298K)$ for the four species are reevaluated based on 35 published chemical kinetic models for $C_2$-$C_5$ hydrocarbons, following the method proposed in [35]. Thereafter, the NASA polynomials for the 24 species are modified based on the reevaluated $\Delta h_f(298K)$, $s^0(298K)$, and $c_p$. The reevaluated thermochemical parameters and 35 published mechanisms are summarized in SMM3, while the modification of thermochemical parameters via the NASA polynomials is briefly discussed herein.

For heat capacity, $c_p$, the NASA coefficients $a_1$-$a_7$ are revised via Eq. 8-10:

$$a_{i,j}^{new} = a_{i,j}^{ori} * \frac{c_p^{mean}}{c_p^{ori}}, j = 1,2,\ldots,5 \tag{8}$$

$$a_{i,6}^{new} = T(1 - \frac{c_p^{mean}}{c_p^{ori}})(a_{i,1}^{ori} + \frac{1}{2}a_{i,2}^{ori}T + \frac{1}{3}a_{i,3}^{ori} + \frac{1}{4}a_{i,4}^{ori}T^3 + \frac{1}{5}a_{i,3}^{ori}T^4) + a_{i,6}^{ori} \tag{9}$$

$$a_{i,7}^{new} = (1 - \frac{c_p^{mean}}{c_p^{ori}})(a_{i,1}^{ori} \ln T + a_{i,2}^{ori}T + \frac{1}{2}a_{i,3}^{ori}T^2 + \frac{1}{3}a_{i,4}^{ori}T^3 + \frac{1}{4}a_{i,5}^{ori}T^4) + a_{i,7}^{ori} \tag{10}$$

For enthalpy, $\Delta h_f(298K)$, the coefficients $a_6$ is revised via Eq. 11:

$$\overline{a_{i,6}^{new}} = a_{i,6}^{new} + \frac{(\Delta h_f(298K)^{mean} - \Delta h_f(298K)^{ori})}{R} \tag{11}$$

For entropy, $s^0(298K)$, the coefficients $a_7$ is revised via Eq. 12:

$$\overline{a_{i,7}^{new}} = a_{i,7}^{new} + \frac{(s^0(298K)^{mean} - s^0(298K)^{ori})}{R} \tag{12}$$

where $c_p^{mean}$, $\Delta h_f(298K)^{mean}$, $s^0(298K)^{mean}$ are the mean values of $\Delta h_f(298K)$, $s^0(298K)$, and $c_p$, respectively. $a_{i,j}^{new}$, $\overline{a_{i,6}^{new}}$, $\overline{a_{i,7}^{new}}$ are the revised NASA polynomial coefficients.

## 3. Results and discussion

*3.1 Reevaluation of the rate parameters*

Figure 2 illustrates the reevaluated rate coefficients for R80: $HO_2+O=OH+O_2$, R83: $HO_2+HO_2=H_2O_2+O_2$, R420 $CH_3CO+O_2=CH_3CO_3$, R212: $CH_2O+CH_3O=HCO+CH_3OH$, and R407: $CH_3CHO+CH_3O_2=CH_3CO+CH_3O_2H$, along with the rate coefficients adopted in Tao et al. [13] and C3MechV3.3 [36], as well as the direct and theoretical data from literature. The results for all other reactions are summarized in SMM4. For R80, as shown in Fig. 2a, Tao et al. [13] and C3MechV3.3 [36] suggested that the rate coefficient is constant over 300-2500 K. However, based on the experimental measurements from Keyser [37] and Peeters [38], the rate coefficient exhibits moderate and non-monotonic temperature dependence. This temperature dependence is better captured by the reevaluated rate coefficient from this study. For R83 (Fig. 2b), the single Arrhenius expression shows poor performance in replicating the data points in the high-temperature region. As such, the rate coefficients are refitted using double Arrhenius expression (i.e., six parameters), where much better agreement with the experimental data is achieved. It can be seen from Fig. 2b that the reevaluated rate coefficient differs slightly from those adopted in Tao et al. [13] and C3MechV3.3 [36], with smaller curvature on the high temperature end. For R420 (Fig. 2c), the reevaluated rate coefficient differs considerably from the recommendation from Tao et al. [13] and C3MechV3.3 [36], both quantitatively and qualitatively. The reevaluated rate coefficient clearly achieves a better representation of the experimental data. In fact, the trend observed with R420 has been found to be quite common among the 79 reactions analyzed, showing that it is necessary to reevaluate $k$ for all important elementary reactions based on direct experiments and theoretical values from quantum chemistry computations. Additionally, for R407 (Fig. 2d) and R212 (Fig. 2e), as well as several other reactions, the rate data is very limited, and only the theoretical data that were recently calculated by Tang et al. [1] can be found. Nevertheless, obvious from Fig. 2d and 2e is the discrepancies between the calculated rate by Tang et al. [1] and those recommended by Tao et al. [13] and C3MechV3.3 [36]. A better agreement is achieved by the refitted rate coefficients, particularly at high temperatures that are more important to combustion modeling.

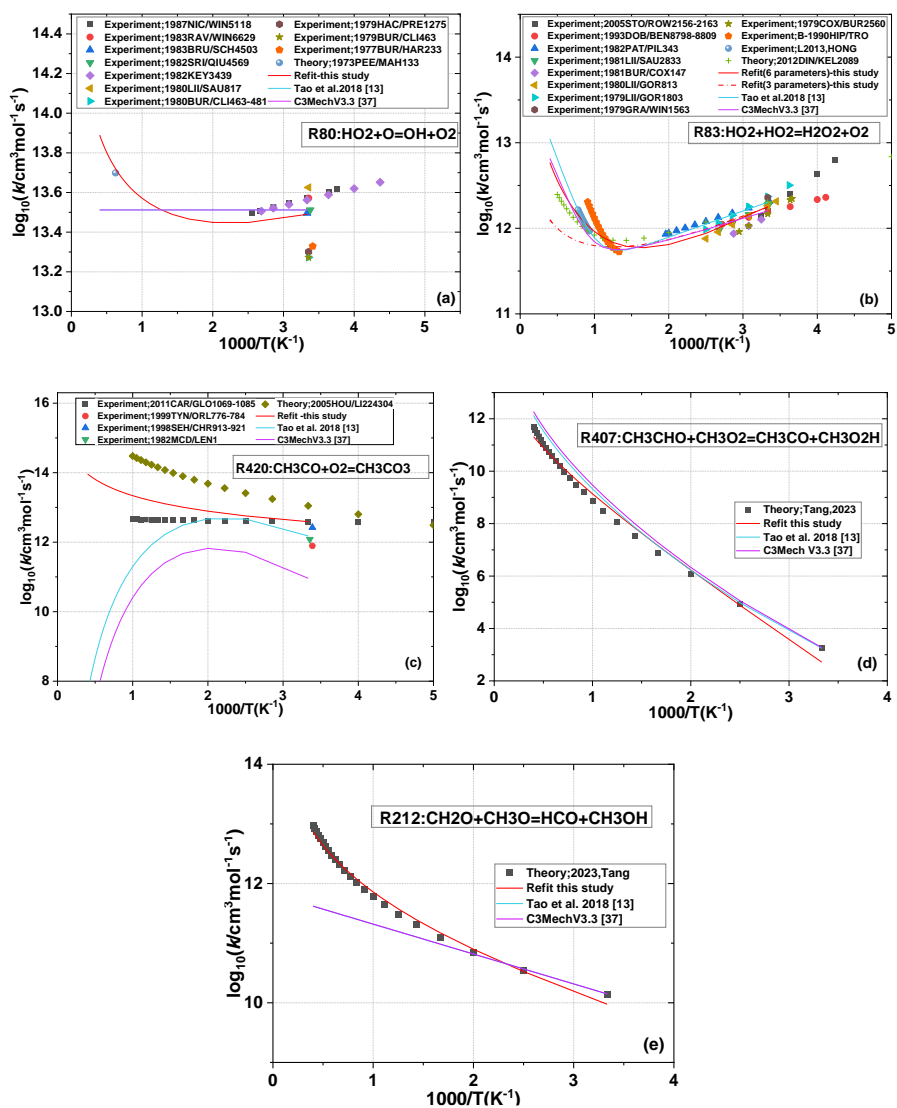

Figure 2. The reevaluated rate coefficients for (a)R80, (b)R83, (c)R420, (d)R407, and (e)R212 (thick red line), along with the values adopted by Tao et al.[13] (thick blue line), C3MechV3.3[36] (thick purple line), and all data used for refitting (symbols).

*3.2 Reevaluation of the thermochemical parameters*

Table 1 summarizes the thermochemical properties ($c_p$(298K), $\Delta h_f$(298K), and $s^0$(298K)) of the 24 selected species computed from Tao et al. [13] and reevaluated in this study. It can be seen in Table 1 that there are significant differences in the heat capacity and entropy of $CH_3O$, the enthalpy and heat capacity of $CH_3O_2$, the heat capacity of $CH_3O_2H$, and the enthalpy of $CH_2CHO$, while the thermochemical properties of other species show relatively small changes.

Table 1. Species selected for optimization, the thermochemical properties parameters ($c_p$(298K), $\Delta h_f$(298K), and $s^0$(298K)) in the initial mechanism and the reevaluated values of the parameters.

| Species | $\Delta h_f$(298K)/ kcal/mol·K | | | $c_p$(298K)/ cal/mol·K | | | $s^0$(298K)/ cal/mol·K | | |
|---|---|---|---|---|---|---|---|---|---|
| | Tao et al. | This study | absolute change | Tao et al. | This study | absolute change | Tao et al. | This study | absolute change |
| $CH_3OH$ | -48.04 | -48.02 | 0.02 | 10.26 | 10.33 | 0.07 | 57.52 | 57.41 | 0.11 |
| $H_2O_2$ | -32.48 | -32.36 | 0.12 | 10.14 | 10.21 | 0.07 | 56.06 | 55.94 | 0.12 |
| $CH_3$ | 35.06 | 35.01 | 0.05 | 9.18 | 9.20 | 0.02 | 46.37 | 46.37 | 0 |
| $CH_4$ | -17.83 | -17.81 | 0.02 | 8.53 | 8.51 | 0.02 | 44.54 | 44.53 | 0.01 |
| HOCHO | -90.48 | -90.46 | 0.02 | 9.87 | 10.14 | 0.27 | 59.07 | 59.17 | 0.1 |
| $HO_2$ | 2.94 | 2.94 | 0 | 8.34 | 8.34 | 0 | 54.76 | 54.76 | 0 |
| $CH_3CHO$ | -39.72 | -39.57 | 0.15 | 13.22 | 13.16 | 0.06 | 63.08 | 63.08 | 0 |
| $CH_3O$ | 5.05 | 5.15 | 0.1 | 10.23 | 9.82 | 0.41 | 54.91 | 55.65 | 0.74 |
| $CH_3O_2$ | 2.74 | 3.12 | 0.38 | 11.62 | 12.42 | 0.8 | 64.36 | 64.43 | 0.07 |
| $CH_3CO$ | -2.46 | -2.40 | 0.06 | 12.14 | 12.22 | 0.08 | 63.90 | 63.90 | 0 |
| $CH_2CO$ | -11.52 | -11.57 | 0.05 | 12.19 | 12.18 | 0.01 | 57.66 | 57.79 | 0.13 |
| $CH_3O_2H$ | -30.39 | -30.52 | 0.13 | 14.16 | 15.79 | 1.63 | 65.97 | 65.81 | 0.16 |
| $CH_3OCH_3$ | -43.84 | -43.98 | 0.14 | 15.39 | 15.56 | 0.17 | 63.90 | 63.85 | 0.05 |
| OH | 8.92 | 8.91 | 0.01 | 7.14 | 7.09 | 0.05 | 43.91 | 43.91 | 0 |
| $C_2H_5$ | 28.92 | 28.44 | 0.48 | 59.08 | 59.17 | 0.09 | 12.01 | 11.92 | 0.09 |
| $CH_2CHO$ | 3.05 | 3.88 | 0.83 | 12.99 | 12.97 | 0.02 | 63.14 | 62.82 | 0.32 |
| HCO | 10.11 | 10.11 | 0 | 8.29 | 8.30 | 0.01 | 53.60 | 53.62 | 0.02 |
| CO | -26.42 | -26.42 | 0 | 6.96 | 6.95 | 0.01 | 47.24 | 47.24 | 0 |
| H | 52.10 | 52.10 | 0 | 4.97 | 4.97 | 0 | 27.42 | 27.40 | 0.02 |
| $H_2O$ | -57.79 | -57.80 | 0.01 | 8.03 | 8.02 | 0.01 | 45.13 | 45.12 | 0.01 |
| $CH_3CO_3$ | -42.35 | -42.36 | 0.01 | 19.91 | 19.85 | 0.06 | 77.73 | 77.49 | 0.24 |
| $HOCH_2O$ | -42.16 | -42.02 | 0.14 | 13.04 | 13.04 | 0 | 66.11 | 65.93 | 0.18 |
| $CH_3CO_3H$ | -80.49 | -80.49 | 0 | 20.48 | 20.48 | 0 | 77.24 | 77.20 | 0.04 |
| $O_2CH_2CHO$ | -21.01 | -21.01 | 0 | 17.28 | 17.28 | 0 | 79.98 | 79.98 | 0 |

*3.3 Model validation and comparison*

The reevaluated kinetic and thermodynamic properties based on direct experiments and theoretical calculations for the 79 reactions and 24 species, respectively, are incorporated into the original model (i.e., that proposed by Tao et al. [13]), which is further used to simulate the available fundamental combustion experiments. The updated model is included in SMM5&6. To assess the predictive capability of the reevaluated model, available fundamental experiments from the literature are adopted for model validation and comparisons with other acetaldehyde chemistry models (e.g., Dagaut et al. [8] and Tao et al. [13]). As summarized in Table 2 and 3, the experiments cover 123 shock tube ignition data over T=1250-1750K, P=2.03-5atm and Φ=0.5-2.5), and 385 JSR species concentration data over T=460-1230K, P=0.934-10atm and Φ=0.43-4. The wide coverage of the thermodynamic and fuel loading conditions should be sufficient to test the reevaluated model.

It is important to note that a major objective of this study is to determine the uncertainty domain for acetaldehyde thermochemistry for the optimization work that is currently in progress in the authors group. Due to the limited experimental and theoretical data on thermodynamic properties for the species identified in this study, it is deemed appropriate to use the recently published mechanisms for this purpose, though that the mechanisms published later might contain better, revised thermodynamics data for the identified species in Table 1. Simulations are first conducted to evaluate the impact of the updated kinetic parameters, the updated thermodynamic parameters, and both on the model performance, with the results summarized in Figs. 3 and 4. It is clear from Figs. 3 and 4 that the change in model performance is dictated by the updates to the kinetic parameters, while the updated thermodynamic parameters exhibit negligible influences on model precisions.

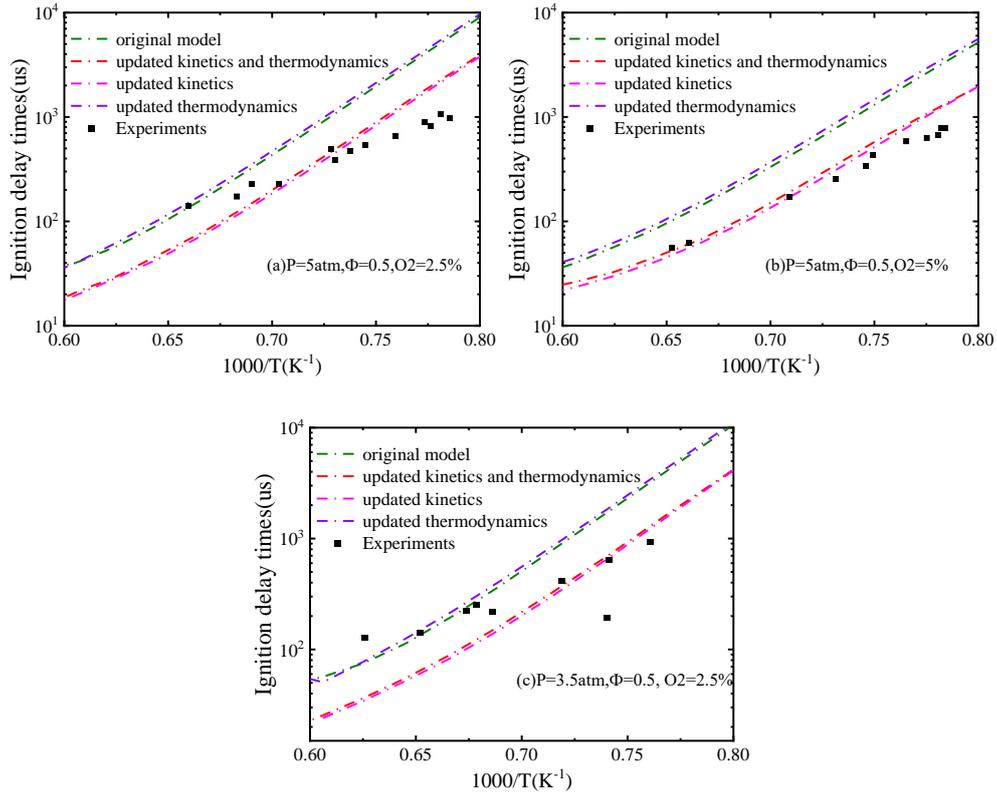

Figure 3. Comparison between measured (symbols) and simulated (lines) ignition delay times for CH3CHO-$O_2$-$N_2$ mixtures in a shock tube. (a) p = 5.0atm; Φ = 0.5; 0.5%CH3CHO / 2.5%$O_2$ / 97%$N_2$. (b) p = 5.0atm; Φ=0.5; 1%CH$_3$CHO / 5%$O_2$ / 94%$N_2$. (c) p = 3.5atm; Φ = 0.5; 0.5%CH$_3$CHO / 2.5%$O_2$ / 97%$N_2$. Experimental data are taken from [8].

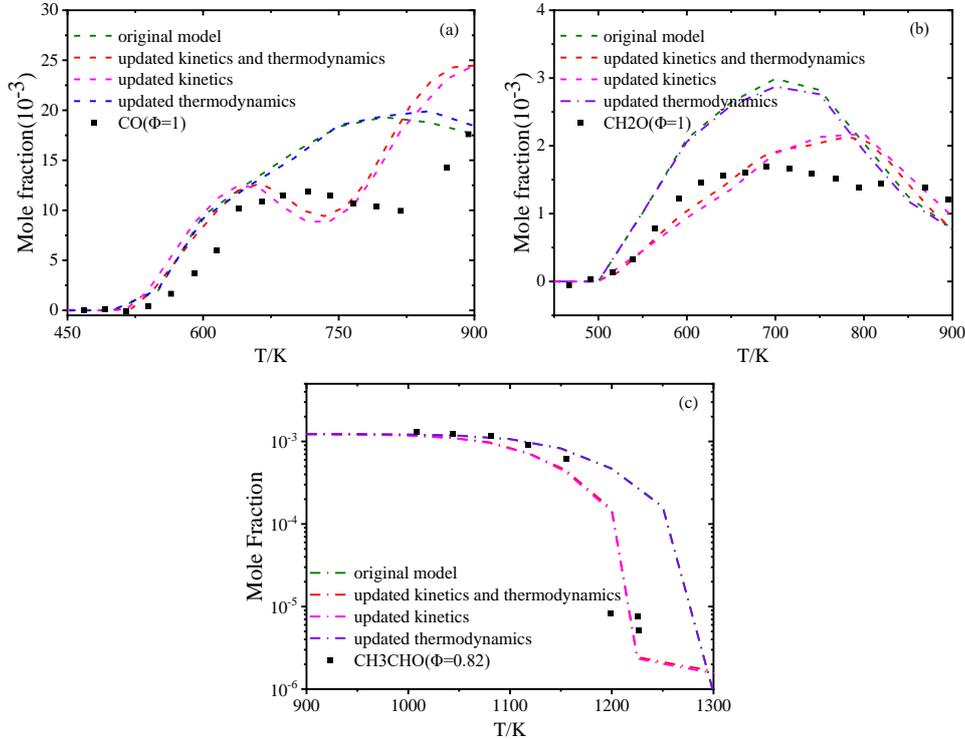

Figure 4. Comparison between measured (symbols) and simulated (lines) species profiles during the oxidation of CH3CHO in a JSR. (a) p = 0.947atm; Φ = 1.0; 2%CH3CHO / 5%$O_2$ / 93%$N_2$. (b) p = 0.947atm; Φ = 1.0; 2%CH3CHO / 5%$O_2$ / 93%$N_2$. (c) p = 1.0atm; Φ=0.82; 0.123%CH3CHO / 0.375%$O_2$ / 99.502%$N_2$. Experimental data are taken from [8].

Figures 5-7 show the simulated results and the experimental measurements for several conditions, while the validation and comparison results at other conditions are included in SMM1. It can be seen from Fig. 5 that, with incorporating the reevaluated rate and thermodynamic parameters, the model becomes more reactive, leading to smaller simulated ignition delay times at all conditions investigated, with stronger change observed at P = 3.5 atm. The temperature dependence of autoignition reactivity seems to be unaffected by the model updates. Compared with the experiments, it is clear that these updates lead to a better agreement between the model and experiments, particularly at temperatures between 1300 and 1400 K. Overall, the updated model performs roughly the same as the model from Dagaut et al. [8]. The difference between the updated model and the Dagaut model is more evident at P = 3.5 atm, where the updated model tends to overpredict the ignition reactivity, while the Dagaut model tends to underpredict the ignition reactivity.

Figure 6 summarizes the simulated and measured JSR species profiles under a fuel-lean condition (i.e., Φ = 0.82) for several important species during acetaldehyde oxidation. It is also clear that the updated model becomes more reactive than the original model, predicting earlier onset of acetaldehyde consumption (c.f., Fig. 6b). Similar to the trends observed in Fig. 5, incorporating the updated model parameters also lead to a better agreement of the model with the fuel-lean JSR experiments for all the species, particularly acetaldehyde (Fig. 6b) and formaldehyde (Fig. 6c). The updated model seems to also outperform the Dagaut model in capturing the species profiles of acetaldehyde (Fig. 6b), formaldehyde (Fig. 6c) and carbon monoxide (Fig. 6f), particularly at temperatures higher than 1150 K. At fuel-rich conditions and low temperatures (750 – 900 K), as shown in Fig. 7, the updated model overestimated the acetaldehyde reactivity, showing significantly higher acetaldehyde consumption rate (Fig. 7b), hence an over predicted carbon monoxide mole fraction, as compared to the experiments. This is due to the increased contributions from the acetaldehyde consuming reactions at this condition, which is shown in Fig. S14 in SMM1.

Table 2. Summary of shock tube data used for model validation and comparison.

| P(atm) | T(K) | Φ | Mole fraction | | | No. of data points | Ref |
|---|---|---|---|---|---|---|---|
| | | | $CH_3CHO$ | $O_2$ | $N_2$ | | |
| 3.45 | 1295-1538 | 0.5 | 0.005 | 0.025 | 0.97 | 6 | |
| 3.45 | 1368-1487 | 1.0 | 0.0086 | 0.0214 | 0.97 | 7 | [10] |
| 3.45 | 1337-1579 | 1.5 | 0.0112 | 0.0188 | 0.97 | 10 | |
| 2.03 | 1275-1619 | 0.5 | 0.01 | 0.05 | 0.94 | 17 | |
| 2.06 | 1275-1619 | 1.0 | 0.02 | 0.05 | 0.93 | 17 | [13] |
| 2.32 | 1399-1704 | 2.5 | 0.02 | 0.02 | 0.96 | 9 | |
| 3.50 | 1315-1598 | 0.5 | 0.005 | 0.025 | 0.97 | 9 | |
| 3.50 | 1365-1735 | 2.0 | 0.01 | 0.0125 | 0.9775 | 12 | |
| 5.00 | 1272-1516 | 0.5 | 0.005 | 0.025 | 0.97 | 13 | [8] |
| 5.00 | 1275-1532 | 0.5 | 0.01 | 0.05 | 0.94 | 11 | |
| 5.00 | 1250-1477 | 1.0 | 0.01 | 0.025 | 0.965 | 12 | |

Table 3. Summary of jet-stirred reactor data used for model validation and comparison.

| Residence time (s) | P(atm) | T(K) | Φ | Mole fraction | | | Data points | Ref |
|---|---|---|---|---|---|---|---|---|
| | | | | $CH_3CHO$ | $O_2$ | $N_2$ | | |
| 0.8 | 10 | 929-1085 | 0.43 | 0.00129 | 0.0075 | 0.99121 | 35 | |
| 0.08 | 1 | 1008-1227 | 0.82 | 0.00123 | 0.00375 | 0.99502 | 44 | [8] |
| 0.8 | 10 | 959-1176 | 1 | 0.0015 | 0.00375 | 0.99475 | 45 | |
| 0.08 | 1 | 1057-1167 | 1.61 | 0.00197 | 0.00186 | 0.99617 | 51 | |
| 3-6 | 0.934 | 461-816 | 0.5 | 0.02 | 0.1 | 0.88 | 49 | |
| 3-6 | 0.947 | 490-894 | 1 | 0.02 | 0.05 | 0.93 | 80 | [12] |
| 3-6 | 0.947 | 560-890 | 4 | 0.02 | 0.0125 | 0.9675 | 81 | |

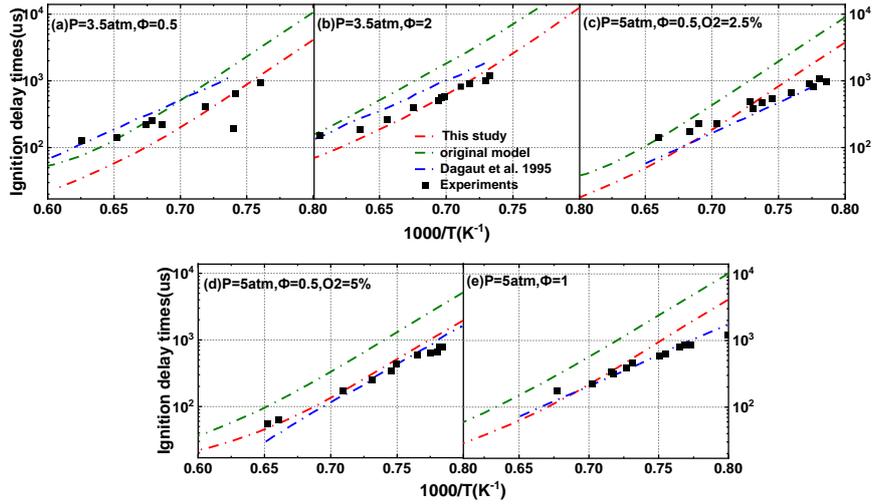

Figure 5. Comparison between measured (symbols) and simulated (lines) ignition delay times for $CH_3CHO$-$O_2$-$N_2$ mixtures in a shock tube. Experiment data are taken from [8].

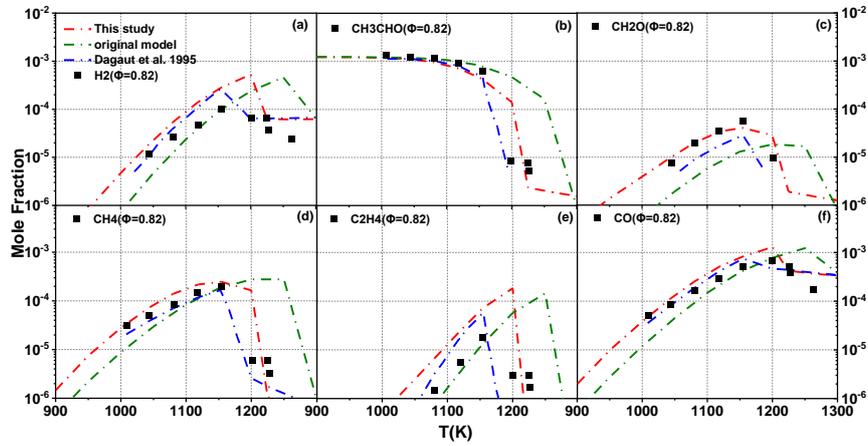

Figure 6. Comparison between measured (symbols) and simulated (lines) species profiles during the oxidation of $CH_3CHO$ in a JSR at P = 1atm. Experimental data are taken from [8].

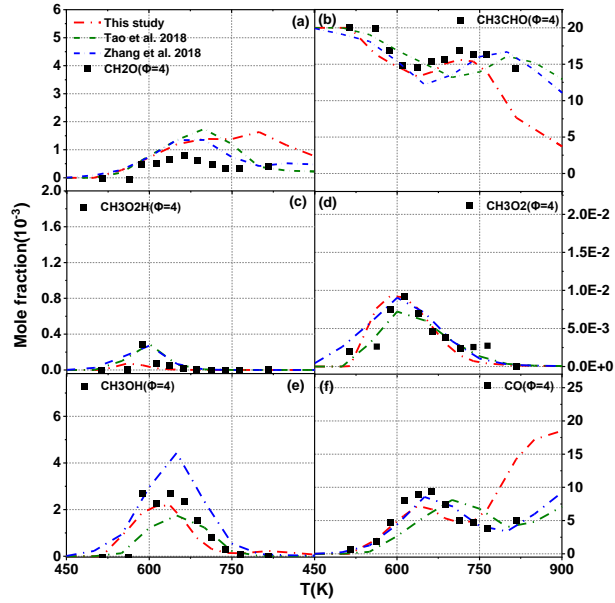

Figure 7. Comparison between measured (symbols) and simulated (lines) species profiles during the oxidation of $CH_3CHO$ in a JSR at P = 0.947atm. Experimental data are taken from [12].

The overall model performance is further evaluated by computing the $L^2$ error norms $F$ between the model predictions and all fundamental experiments using

$$F = \sqrt{\frac{1}{n}\sum_{i=1}^{n}(\frac{\eta_i - \eta_i^{exp}}{\sigma_i^{exp}})^2} \quad (13)$$

where $n$ is the number of experimental measurements, $\eta_i$ and $\eta_i^{exp}$ are the model predictions and the experimental measurements, respectively, and $\sigma_i^{exp}$ is the measured uncertainty of the ST and JSR experiments, which is inferred from the error bars presented in the literature or nominally defined as ±15% when error bars are not available. The results are shown in Fig. 8a. It is clear from Fig. 8a that the updated model is comparable to the other models (including the models from Tao et al. [13], Zhang et al. [12], Daugaut et al. [8], and Mével et al. [10]) at below 1200 K, while exhibiting much better agreement with experiments at higher temperatures and the highest pressure conditions (e.g., 10 atm). Fig. 8b further shows the absolute difference between the update model and experiments at various temperatures (400 – 1800 K), pressures (0.9 – 10 atm) and equivalence ratios (0.43 – 4.00), where a better agreement is observed at higher temperatures, which is consistent with the trends observed in Fig. 8a. Also seen from Fig. 8b is the somewhat better model performance at higher pressure conditions than at low pressure conditions.

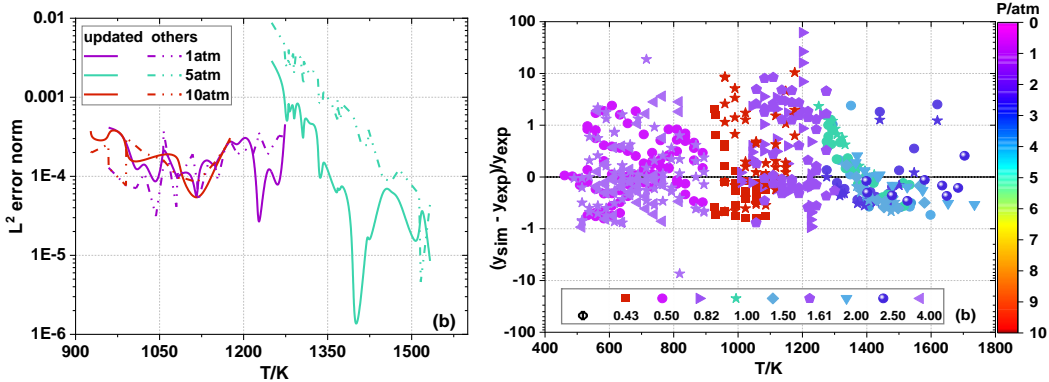

Figure 8. (a) $L^2$ error norms of the updated model (solid lines) and other models (dash-dot lines), and (b) absolute difference (in percentage) between predictions using the updated model and experiments (Symbol represents different equivalence ratios, and color indicates pressure conditions). Others refer to the simulations results using Tao et al. [13], Zhang et al. [12], Daugaut et al. [8], and Mével et al. [10] mechanisms here.

It is interesting to see that although the updated model is not adjusted against these fundamental experiments (i.e., all the important reactions and species are reevaluated based on direct experiments and theoretical calculations for rate and thermodynamic parameters), it yields a satisfactory performance in replicating all the fundamental experiments, which is overall equivalent to the other models. At some conditions, the model even outperforms (e.g., Fig. 5) the other models that have already been calibrated based on the reported experiments. This highlights the validity of physics-based reevaluation of the model parameters using direct experiments and high-level theoretical computations.

*3.4 Sensitivity and flux analyses*

To explore the underlying kinetics that lead to the changes in model performance, sensitivity analysis is performed at T=750 and 1150K for the ignition delay time of a stoichiometric acetaldehyde mixture at P = 10 bar, using both the original model from Tao et al. [13] and the updated model. The sensitivity analysis results are shown in Figure 9, covering the top 20 most sensitive reactions. Negative sensitivity coefficients indicate that the reaction promotes reactivity, while positive coefficients indicate an inhibiting effect. At T=750K, the sensitivity coefficients computed by the original model and the updated model are significantly different, with almost all the promoting and inhibiting reactions exhibiting stronger contributions to autoignition reactivity. The promoting effect from the top promoting reaction, i.e., $CH_3CO+O_2 = CH_3CO_3$, is enhanced by approximately 70%. This is expected as the rate coefficient of this reaction has been increased considerably in the updated model, as shown in Fig. 2c. The enhanced promoting effect counteracts the enhanced inhibiting effect, which might result in an overall negligible impact on model reactivity at this temperature. This is confirmed in Fig. 7, where the simulated species mole fraction for acetaldehyde and carbon monoxide are nearly identical between the original and updated model. At T=1150K, as shown in Figure 9b, most reactions have similar sensitivities between the updated and original models. However, the sensitivity of the most inhibiting reaction in the original model, namely R99 ($CH_3+HO_2=CH_4+O_2$), is reduced by more than half, and the sensitivity of the most promoting reaction ($CH_3CHO+HO_2=CH_3CO+H_2O_2$) is increased by around 100%. These changes will lead to an increased model reactivity. This, again, agrees well with the trends seen in Fig. 5-7 and Fig. S1-S12 in SMM1 where the updated model exhibits higher reactivity at 1150 K and high temperatures.

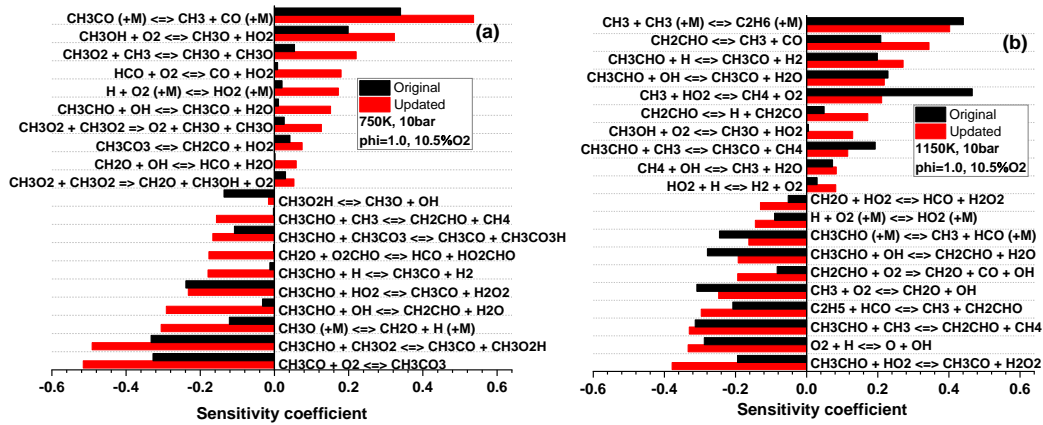

Figure 9. Sensitivity analysis on ignition delay time for Original and Updated model at the diluted/stoichiometric condition, P=10 bar and different temperatures; (a)T=750K and (b)1150K.

Flux analysis is further conducted at the same conditions as in Fig. 9 at the timing of 1% $CH_3CHO$ consumption. The results at T=750 and 1150 K are shown in Fig. 10 and Fig. S13 in SMM1, respectively. As shown in Fig. 10, significantly different branching ratios are observed at 750 K for the conversions from $CH_3CHO$ to $CH_3CO$ and $CH_2CHO$, as well as from $CH_2O$ to HCO. Specifically, with the updated rates, the contribution of H-atom abstraction by OH radicals to $CH_3HO$ consumption increases considerably, while that by $CH_3O_2$ radicals is suppressed. This is consistent with the SA results, where the sensitivity of $CH_3CHO$+OH pathways is increased by more than 10 times. There is also a clear shift toward the formation of $CH_2CHO$ with the updated model (i.e., <2.5% and 27% $CH_3CHO$ is consumed to form $CH_2CHO$ in the original and updated model, respectively). These $CH_2CHO$ producing pathways promote reactivity, as can be seen in Fig. 9a. An increased flux via these pathways will lead to enhanced reactivity, supporting the shifted model reactivity as observed in Fig. S1-S12 in SMM1. Additionally, a prominent rise and decrease is seen in the $CH_2O$ consumption pathways by OH and $CH_3O_2$, respectively. Similar trends to those observed at 750 K are found at 1150 K (Fig. S13), with smaller differences seen between the original and updated models.

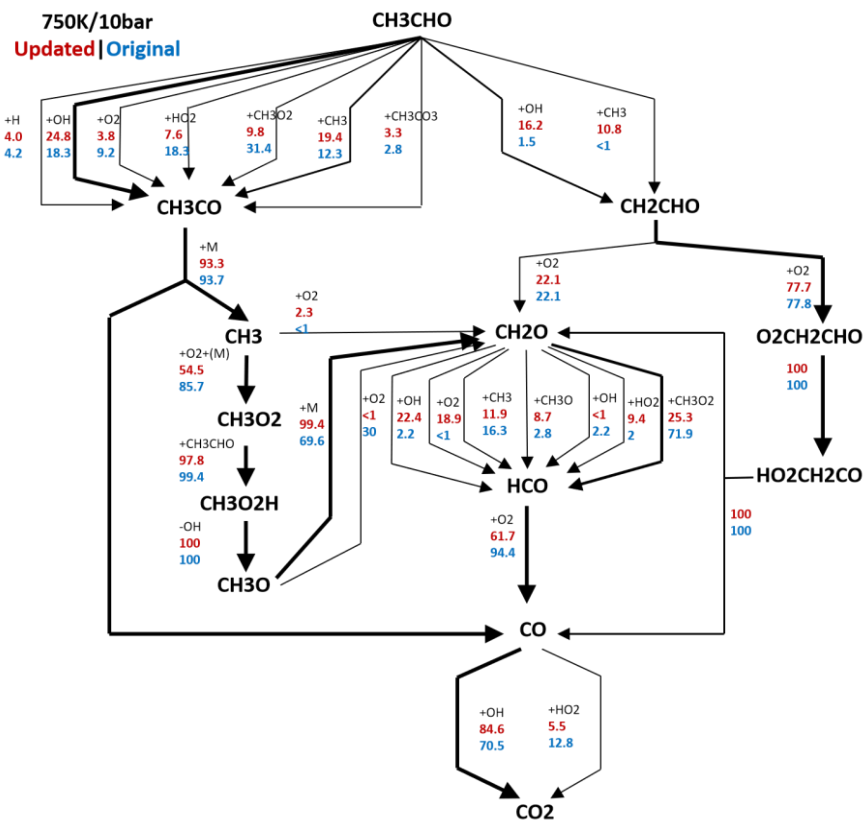

Figure 10. Flux and rate of production (ROP) analyses for updated (red numbers) and original (blue numbers) at 750K, 10 bar.

## 4. Uncertainty analysis

As can been seen from Figs. 5-7, the disagreements between the updated model and the experiments are still significant, though slightly improved at high temperatures. These disagreements can be addressed by further optimizing the model parameters within the model uncertainty. Doing so, the model uncertainty needs to be defined with addressing two important questions: (a) is the parameter uncertainty rigorously defined based on known information? and (b) is the nominal parameter value a statistical representation of its uncertainty domain? Without properly addressing any of these information, the model uncertainty will be falsely conditioned, and thus empty solution will be returned when enough constraints have been applied during model optimization (e.g., data inconsistency) [39]. Fortunately, these two questions can now be addressed based on the updated chemistry model developed in this study, as the model parameters are evaluated purely based on direct experiments and quantum chemistry computations, which is a statistical representation of the respective parameter uncertainty domain. To this end, the uncertainties for the 79 reactions and the 24 species are evaluated in the following.

Firstly, the temperature-dependent uncertainty of the identified reactions is determined following the method proposed by Nagy and co-workers [40], while basing on the reevaluated reaction rates rather than the mean rate coefficients as adopted in [40]. The temperature-dependent uncertainty factor $f(T)$ are determined from the uncertainty bands $log_{10}(\kappa(T, \mathbf{x}^*)) \pm f(T)$, which are derived from the limiting values of the acceptable rate coefficients based on direct experiments, quantum chemistry computations, as well as review studies. The uncertainty factor corresponds to $3\sigma$ deviations from the recommended values on a logarithmic scale $\sigma_\kappa(T) = f(T) \cdot (ln10)/3$.

The relation between the variance of $\kappa(T)$ and the elements of the covariance matrix of the Arrhenius parameters is further obtained via the method proposed by Nagy et al. [40]:

$$\sigma_\kappa(T) = \sqrt{\Theta^T \Sigma_p \Theta} = \sqrt{\sigma_\alpha^2 + \sigma_n^2 ln^2 T + \sigma_\varepsilon^2 T^{-2} + 2r_{\alpha n}\sigma_\alpha\sigma_n lnT - 2r_{\alpha\varepsilon}\sigma_\alpha\sigma_\varepsilon T^{-1} - 2r_{n\varepsilon}\sigma_n\sigma_\varepsilon T^{-1}lnT} \quad (14)$$

where $\Sigma_p$ is the covariance matrix of parameters $(\alpha, n, \varepsilon)$, which ($\sigma_\alpha^2, \sigma_n^2, \sigma_\varepsilon^2$) are variances, and ($r_{\alpha n}, r_{\alpha\varepsilon}, r_{n\varepsilon}$) are correlations. The covariance matrix $\Sigma_p$ can be used to restore $f(T)$, which also defines the prior uncertainty domain of Arrhenius parameters.

For reactions involving third-bodies, additional treatment is needed to reconcile the data sources reported for different third bodies. In reevaluating the uncertainty for this type of reaction, the third-body collision efficiencies are pre-determined from literatures (see SMM4) with that of $N_2$ set at unity and others set at different values relative to unity. An example is shown in Fig. 11 for R85: $H+O_2(+M)=HO_2(+M)$. It is clear from Fig. 11 that the reevaluated rate coefficient from this study represents all the data better than the others, particularly in the high-temperature regime. The temperature-dependent uncertainty factors determined based on the reevaluated rate coefficient for R85 are shown in Fig. 11b, while the results for other reactions are shown in SMM4. Also seen in Fig. 11 is the necessity of rate coefficient reevaluation for determining uncertainty factors. Defined based on the reevaluated rate coefficient, the uncertainty bands, as shown in Fig. 11b, are able to enclose almost all reported experimental and theoretical values. However, the rate coefficients from Tao et al. and C3MechV3.3 fall outside the physical uncertainty bands at high temperatures. If uncertainty bands are determined based on the rate coefficients from Tao et al. and C3MechV3.3, the uncertainty bands will certainly shift away from the uncertainty domain, undermining the fidelity of subsequent model optimization. This process is repeated for all 79 reactions, with the results for other reactions summarized in SMM4.

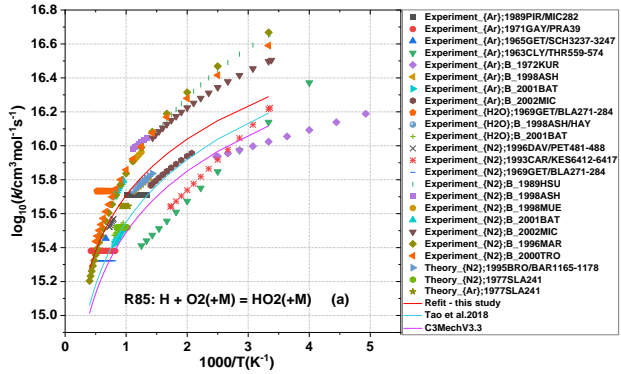

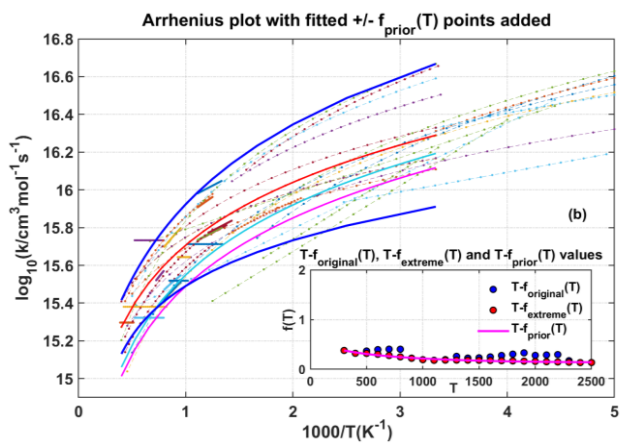

Figure 11. (a) Reevaluated rate coefficients for R85 (symbols are labelled in the legend as *type of data {bath gases}: NIST ID*), (b) Upper and lower uncertainty limits (thick blue line) for R85, with inset illustrating temperature-dependent uncertainty factors: $f_{original}$ (blue dots) and $f_{extreme}$ (red dots), and $f_{prior}$ (purple line).

The temperature-dependent uncertainty factors for each reaction can be obtained by calculating the covariance matrix of the Arrhenius parameters. This information is also beneficial for observing the uncertainty factor $f(T)$ of interested reaction at the specific temperature, which can facilitate future optimization. For some reactions (e.g., R407 and R441 in Fig. 12a), the rate data are very limited, making it difficult to determine the covariance matrix elements of the Arrhenius parameters. Therefore, an uncertainty factor of 2 is defined over the entire temperature range studied. Fig. 12 shows the $f(T)$ calculated from the covariance matrix of the Arrhenius parameters for the 15 most sensitive reactions, which are selected from the sensitivity analysis of kinetic parameters based on the IDT at T= [600, 900, 1500] K, P=80bar, Φ=1. It is obvious from Fig. 12 that most reactions demonstrate higher uncertainties at low temperatures compared to high temperatures, particularly for R171: $CH_3OH+O_2=CH_3O+HO_2$ and R401: $CH_3CHO+HO_2=CH_3CO+H_2O_2$. The differences in uncertainty factors between different reactions are also significant. For instance, the uncertainty factor $f(T)$ of R171 changes between 5.5 and 1.0 in the range of temperature 300-800K, while that for R401 changes between 3.0 and 1.0 in the range of temperature 300-550K.

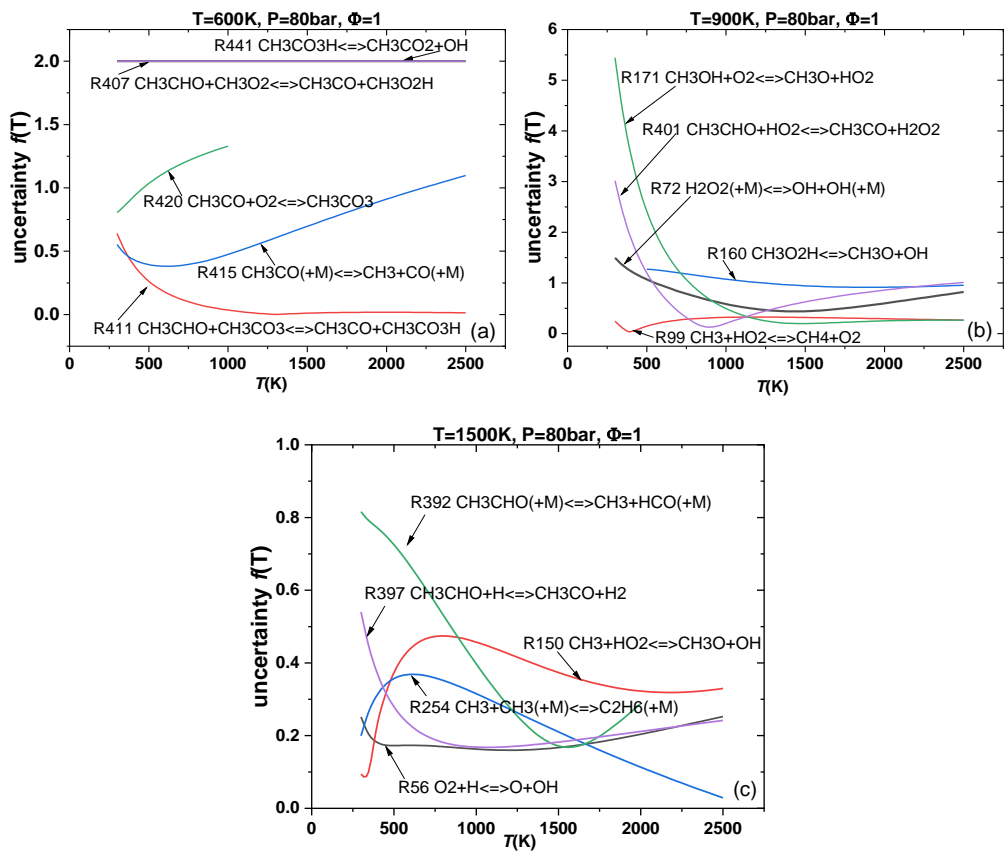

Figure 12. Uncertainty $f(T)$ curves for the investigated reactions. These lines were calculated from the covariance matrix of the Arrhenius parameters.

Figure 13a summarizes the overall deviation of the reevaluated rate coefficients from the original rate coefficients at different temperatures. The 79 reactions are evaluated at each temperature using the original and reevaluated rate parameters, from which the difference is computed and statistically treated with normal distribution. It can be seen from Fig. 13a that the reevaluated rate coefficient can differ from the original rate coefficient by 8 magnitude orders (i.e., at 300 K). Overall, greater differences are seen at temperatures below 1100 K (i.e., the low- and intermediate-temperature regimes). The large differences between the reevaluated and original rate coefficients, again, highlight the necessity of reevaluation based on experimental and theoretical calculation on rate parameters. Figure 13b shows the overall distribution of rate parameter uncertainties at different temperatures, which is computed from the uncertainty factors of the 79 reactions at each temperature. Again, it is obvious from Fig. 13b that the uncertainty space for model optimization is large and that there is a high potential for the model to be optimized, particularly within the low-temperature regime.

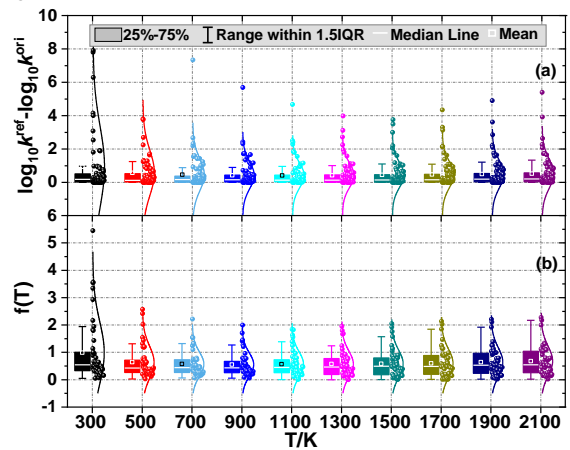

Figure 13. Absolute values and normal distribution of (a) $\log_{10}\kappa^{new}-\log_{10}\kappa^{ori}$ and (b) the uncertainty factors $f(T)$ for the 79 reactions at different temperatures.

Following the above, the temperature-independent uncertainty of the thermodynamic parameters is also investigated, which represents $2\sigma$ deviations from the reevaluated values. Fig. 14 shows the uncertainty of thermochemical properties of 24 key species where the computed values are minus by the respective mean value. Several species (e.g., $CH_3O$, $CH_3O_2$, $CH_3CO$, $CH_2CO$, $CH_3CO_3$) clearly show greater uncertainties than their peers. Again, it is obvious from Fig. 14 that there is a large potential for the thermodynamic parameters to be optimized.

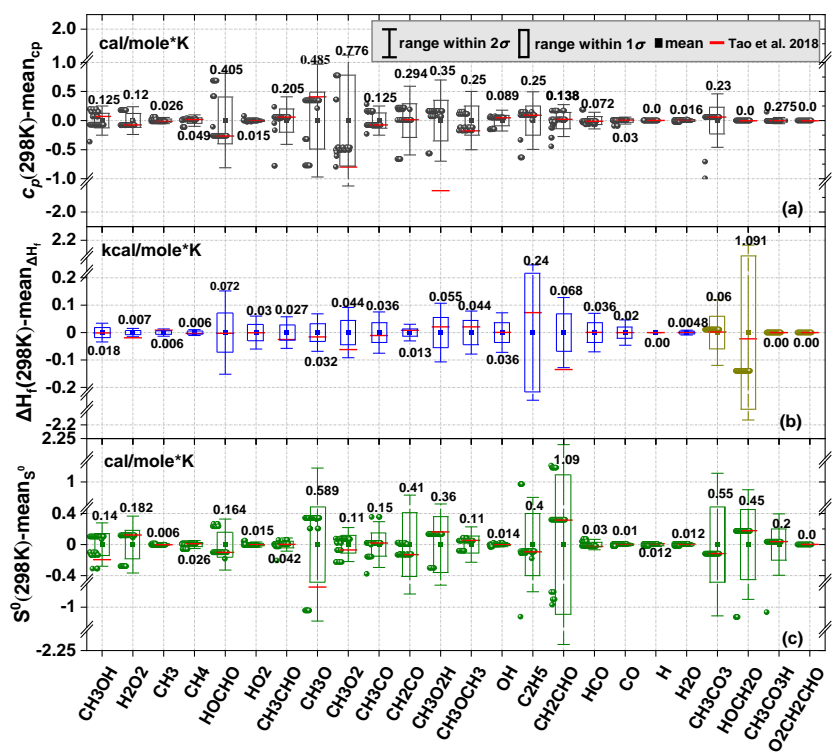

Figure 14. Statistical distribution and uncertainty of thermochemical properties for the 24 species computed from databases and (or) 35 published chemistry models: (a) $c_p$(298K), (b) $\Delta h_f$(298K), and (c) $s^0$(298K). The numbers on each column represent the absolute value of $1\sigma$.

## 5. Conclusion

This study aims to define a new chemical kinetic model of acetaldehyde. 79 elementary reactions and 24 species are identified via sensitivity analyses over the temperature, pressure and fuel-loading conditions that are of interest to the combustion community. The rate parameters of the 79 reactions are reevaluated using over 100,000 direct experiments and quantum chemistry computations from >900 studies, and the thermochemistry ($\Delta h_f$(298K), $s^0$(298K) and $c_p$) are reevaluated based on the ATCT database, the NIST Chemistry WebBook, the TMTD database, and 35 published chemistry models.

The updated rate parameters and thermochemistry are incorporated into a recent acetaldehyde chemistry model, which is further assessed against available fundamental experiments, including 123 ignition delay times and 385 species concentrations. Four published acetaldehyde models are also adopted for comparison. Surprisingly, the updated model, though not calibrated against the fundamental experiments, are equal to or better than the previous models, despite the large differences in kinetic and thermochemical parameters of the key reactions and species, respectively. The new model clearly better reproduces the experiments in the high-temperature regime. Sensitivity and flux analyses are further performed, with considerably changed branching ratios in the updated model for the consuming pathways of $CH_3CHO$ and $CH_2O$. There is a clear shift toward H-atom abstractions by OH radicals for $CH_3CHO$ and $CH_2O$ consumption, with greatly promoted $CH_2CHO$ formation.

The large differences between the updated rate coefficients and the original values adopted in the existing chemistry models reveal the necessity of reassessment of key reactions in the core chemistry completely based on direct experiments and theoretical calculations for rate parameters. Following this, based on these datasets, as well as recommended values from review articles, the temperature-dependent and temperature-independent uncertainties are statistically evaluated for kinetic and thermochemical parameters, respectively. This uncertainty will be used as the initial uncertainty domain for parameters optimization calculation.

Since the model uncertainty is representative of the uncertainties of the rate and thermodynamic parameters and the nominal model can be considered as a statistical representation of its uncertainty, a unified framework for real-time UQ and optimization can be developed with minimized probability of data inconsistency, which will be demonstrated in a forthcoming study from the author's group where the observed discrepancies between the updated model and the fundamental experiments will be minimized.

**CRediT authors Contributions Statement**

**Xinrui REN:** Conceptualization, Methodology, Investigation, Writing – original draft. **Hongqing WU:** Methodology, Investigation. **Ruoyue TANG:** Methodology, Investigation. **Yanqing CUI:** Methodology, Investigation. **Mingrui WANG:** Methodology, Investigation. **Song CHENG:** Conceptualization, Investigation, Writing – review & editing, Funding acquisition, Supervision.

**Declaration of competing interest**

The authors declare that they have no known competing financial interests or personal relationships that could have appeared to influence the work reported in this paper.

**Data availability**

Data will be made available on request.

**Acknowledgements**

The work described in this paper was supported by grants from the Research Grants Council of the Hong Kong Special Administrative Region, China (PolyU P0046985 for ECS project funded in 2023/24 Exercise and P0050998), and by the Natural Science Foundation of Guangdong Province under 2023A1515010976.

**Supplementary materials**

SMM1 and SMM2 list the sensitivity analysis results of the kinetic and thermodynamic parameters, respectively; SMM3 summarizes reevaluation results and the determined uncertainties for the 79 elementary reactions and 24 species; SMM4 includes the model validation and flux analysis results; SMM5 and SMM6 are the updated acetaldehyde model files.